\documentclass[preprint]{revtex4}
\newcommand{\ba}{\begin{eqnarray}}
\newcommand{\ea}{\end{eqnarray}}

\usepackage{graphics}
\begin{document}

\title{ Effect  of electron-nuclear spin interactions 
on electron-spin qubits localized in self-assembled quantum dots}

\author{Seungwon Lee, Paul von Allmen, Fabiano Oyafuso, and Gerhard Klimeck} 
\affiliation{Jet Propulsion Laboratory, California
Institute of Technology, Pasadena, California 91109} 
\author{K. Birgitta Whaley}
\affiliation{Department of Chemistry, University of California, Berkeley, California 94720}
\date{\today}

\begin{abstract} 
The effect of electron-nuclear spin interactions on qubit operations is investigated 
for a qubit represented by the spin of an electron   localized in a self-assembled quantum dot. 
The localized electron wave function is evaluated within the atomistic tight-binding model.
The magnetic field generated by the nuclear spins is estimated in the presence of 
an  inhomogeneous environment  characterized by a random nuclear spin configuration, 
by the dot-size distribution, by alloy disorder, and by interface disorder. 
Due to these inhomogeneities, the magnitude of the nuclear magnetic field varies 
from one qubit to another by the order of 100~G, 100~G, 10~G, and 0.1~G, respectively.  
The fluctuation of the magnetic field causes errors in exchange operations 
due to the inequality of the Zeeman splitting 
between two qubits. We show that the errors can be made lower than the quantum error threshold 
if an exchange energy larger than 0.1~meV is used for the operation.  
\end{abstract}

\maketitle 

\section{Introduction}
Quantum computers (QC) hold the promise of solving problems that would otherwise be beyond the practical range of conventional computers. 
Natural candidates for the fundamental building block of quantum computers (qubit)  
are  the electronic and the nuclear spin, since  they have a well defined Hilbert space and a relatively 
long decoherence time compared to the orbital degrees of freedom. 
Several QC implementations have been proposed based on the use of the spin degree of freedom, such as using the nuclear spins in a molecule\cite{gershenfeld} and in crystal lattices,\cite{yamaguchi} the nuclear spins of donors in Si\cite{kane} and in endohedral fullerenes,\cite{twamley} and the spin of electrons confined in quantum dots\cite{loss-divincenzo, imamoglu} or donors.\cite{vrijen} 
While small-scale quantum computing has been demonstrated with a few qubits, using 
the nuclear spin\cite{chuang-vandersypen, vandersypen-steffen-prl, vandersypen-steffen-nature}  or trapped ions,\cite{gulde} large-scale quantum computing with many qubits used in parallel is yet  to be demonstrated. 

Solid-state spin-based QC architectures are in principle scalable to many qubits. 
However, they are intrinsically inhomogeneous due to defects, impurities, interfaces, etc.
Although an inhomogeneous environment will cause inaccuracy in the qubit operations, fault-tolerant error-correction schemes will compensate for quantum errors if the error occurrence rate is smaller than $10^{-3} - 10^{-4}$ per operation.\cite{steane} 
For spin-based QC architectures, the single-qubit operation typically uses 
the Zeeman coupling to an external magnetic field ($g\mu_B{\bf S}\cdot{\bf B}$),
while the two-qubit operation relies on the exchange interaction between two spins ($J{\bf S}_1\cdot{\bf S}_2$). 
Therefore, if the inhomogeneous environment causes fluctuations in the local magnetic field,
it will lead to errors in the quantum operations.  
For example, a recent study\cite{hu-sousa-sarma} has shown that a ``swap" operation with a magnetic field fluctuation $\Delta B$ yields an error of $(g\mu_B\Delta B/ J)^2$.  
It is thus crucial to examine whether the proposed solid-state spin-based QC implementations are scalable within the quantum error limit, in presence of a realistic inhomogeneous environment.  
As a prototype of this examination, the present paper focuses on the scalability of architectures where the qubit is represented by an excess electron spin localized in a self-assembled InAs quantum dot. 
An array of self-assembled quantum dots is an excellent candidate for a scalable QC architecture because recent advances in the fabrication technology have substantially improved the control of size and location of the nanostructures. 
 
Among the many sources of inhomogeneity in the local magnetic field for an InAs quantum dot embedded in a GaAs buffer, 
the electron-nuclear spin interaction is responsible for the largest magnetic field fluctuation. 
All the nuclei in this nanostructure possess a nonzero magnetic moment, and the number of nuclei interacting with the electron spin 
is in the range $10^{4} - 10^{6}$.  The effective magnetic field generated for each electron by such a large number of nuclear spins 
varies from dot to dot due to the random nuclear spin orientation, the dot-size distribution, alloy disorder, and interface disorder. 
In this paper, we first estimate the fluctuations in the effective nuclear magnetic field resulting from the hyperfine coupling, calculating the electron density using the atomistic tight-binding model and including the fluctuation of nuclear magnetic moments due to the inhomogeneities in the environment. 
The tight-binding model is ideally suited for the description of alloy and interface disorder with atomistic resolution, which enables us to study the microscopic effect of the inhomogeneous environment on the electron densities. 
In a second stage, we evaluate the effect of the resulting fluctuations in the nuclear magnetic field on single qubit and two-qubit operations. 

The paper is organized as follows.
Section~\ref{sec:interaction} describes the treatment of the electron-nuclear spin interaction
within the tight-binding model. Section~\ref{sec:BN} describes the spatial fluctuation of the effective nuclear magnetic field acting on 
the electron spin from one quantum dot to another due to the inhomogeneous environment.  
Section~\ref{sec:operation} discusses the effect of the spatial fluctuation of the nuclear magnetic field on qubit operations. 
Finally, Section~\ref{sec:conclusion} summarizes the results of this work.    

\section{Electron-nuclear spin interaction}\label{sec:interaction}
      
The electron-nuclear spin interaction originates from the coupling of a nuclear magnetic moment to 
the magnetic field generated by an electron magnetic moment (or equivalently 
from the coupling of an electron magnetic moment to the magnetic field generated by a nuclear magnetic moment).
Both the spin and orbital angular momentum of the electron contribute to its magnetic moment.
However, the conduction electron wave function for  InAs self-assembled dots is mostly ($>$90\%) composed of $s$-symmetry 
atomic orbitals and hence the orbital magnetic moment of the electron can be ignored.  
As a result, the remaining electron-nuclear spin interaction is described by the hyperfine Fermi contact interaction:
\ba
\hat{H}_{HF}=\frac{16\pi}{3}\mu_B \mu_N \sum_j g_j (\hat{\bf S}\cdot\hat{\bf I}_j) \delta(\bf{r}-{\bf R}_j),  
\ea 
where $\mu_B$ and $\mu_N$ are the Bohr magneton and the nuclear magneton, and $g_j$ is the $g$ factor of 
the $j$th nuclear spin. $\hat{\bf S}$ and $\hat{\bf I }_j$ are the spin operators for the electron and the $j$th nucleus, and  
${\bf r}$ and ${\bf R}_j$ are the position vectors for the electron and the $j$th nucleus. 
Since the energy of the hyperfine interaction ($<$0.1~meV) is much smaller than the energy spacing between the quantized 
electron levels (about 10--100~meV), the hyperfine Hamiltonian for a given electron level can be approximated with first order perturbation theory as:
\ba
\hat{H}_{HF}&=&\frac{16\pi}{3}\mu_B\mu_N \sum_j g_j |\psi({\bf R}_j)|^2(\hat{\bf S}\cdot\hat{\bf I}_j) \nonumber \\
                       &=&\sum_j A_j  (\hat{\bf S}\cdot\hat{\bf I}_j), \label{eq:HF}
\ea
where $\psi({\bf R}_j)$ is the electron wave function at nuclear site ${\bf R}_j$, and $A_j$ is 
the effective hyperfine coupling constant between the electron and the $j$th nuclear spin.

The coupling constant $A_j$ is proportional to the square of the electron wave function at a nuclear site:
\ba
A_j = \frac{16\pi}{3}\mu_B \mu_N g_j \left|\psi(\bf{R}_j)\right|^2.  \label{eq:Aj}
\ea
Within the tight-binding model, the electron wave function is expressed as a linear combination
of atomic basis orbitals $\phi({\bf r}-{\bf R}_j)$. 
The present tight-binding model includes $sp^3d^5s^*$ basis orbitals.\cite{boykin_strain} 
Therefore, the total electron density at nuclear site ${\bf R}_j$ is given by 
\ba
|\psi({\bf R}_j)|^2 = |\alpha_j \phi_s(0) + \beta_j \phi_{s^*}(0)|^2,
\label{eq:density}
\ea 
where $\alpha_j$ and $\beta_j$ are the tight-binding coefficients for $s$ and $s^*$ orbitals 
centered at site ${\bf R}_j$, respectively. 
In terms of the effective mass approximation, the tight-binding coefficients loosely speaking correspond to the envelope functions while the tight-binding orbitals correspond to the Bloch wave functions. 

\begin{figure}[t]
\scalebox{0.8}{\includegraphics*{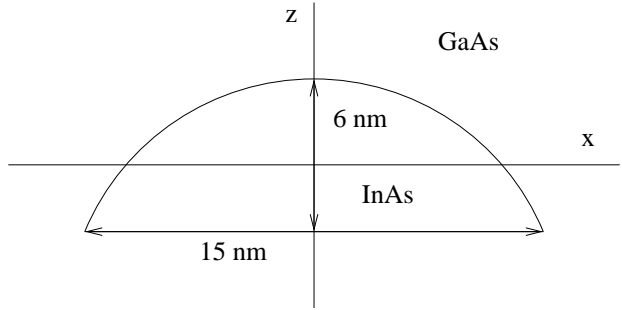}}
\caption{ Geometry of the self-assembled quantum dot modeled in this paper.  
The quantum dot is lens shaped with a base diameter of 15~nm and a height of 6~nm. 
The lines $x$ and $z$ are the lateral and vertical axes along which the spatial distributions of $A_j$ are plotted in Figure~\ref{fig:Aj}.}
\label{fig:dotgeometry}
\end{figure}

The tight-binding coefficients $\alpha_j$ and $\beta_j$ depend on dot geometry, material, 
strain profile, alloy disorder, etc. 
The geometry of a quantum dot grown by Molecular Beam Epitaxy varies widely 
with the growth condition.\cite{moison, kobayashi, mukhametzhanov}
Based on the experimentally achievable geometries,   
we model a lens-shaped self-assembled InAs dot with diameter 15~nm and height 6~nm, as shown in Figure~\ref{fig:dotgeometry}.
Since the dots are embedded in a GaAs matrix, InAs/GaAs self-assembled dots are strongly strained due to the large lattice mismatch of 7\% between InAs and GaAs.  
The equilibrium atomic positions under strain are calculated with an atomistic valence force field model\cite{keating}. 
The strain effect on the electronic structure is captured by adjusting the atomic energy levels with 
a linear correction that is obtained within the L\"owdin renormalization procedure.\cite{lowdin, boykin_strain}   
We also modify the nearest-neighbor coupling parameters for the strained structures according to the generalized version of Harrison's $d^2$ scaling law and the Slater-Koster direction-cosine rules.\cite{harrison, slater-koster}   

In the empirical tight-binding model, $\phi_s(0)$ and $\phi_{s^*}(0)$ are unknown 
because the model determines the Hamiltonian matrix elements without 
introducing the real-space description of the basis orbitals. 
For this work, the densities of the basis orbitals at a nuclear site are determined empirically using measurements of the Overhauser shift of the electron spin resonance.\cite{gueron}  
The details of determining the densities are given in the Appendix. 

\begin{figure}
\scalebox{0.5}{\includegraphics*{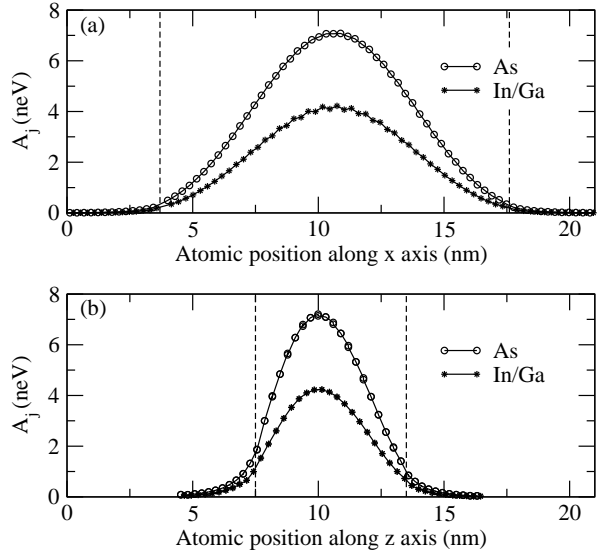}}
\caption{Spatial distributions of the hyperfine coupling coefficient ($A_j$) 
for an InAs quantum dot embedded in a GaAs buffer  
(a) along the $x$ axis and (b) along the $z$ axis of Fig.~\ref{fig:dotgeometry}. 
The coupling coefficient $A_j$ is given by Eq.~(\ref{eq:Aj}) and is proportional to the electron density. 
The dashed lines indicate the interface between the InAs dot and the GaAs buffer. }
\label{fig:Aj}
\end{figure}

With the resulting $\alpha_j$, $\beta_j$, $\psi_s(0)$, and $\psi_{s^*}(0)$, $A_j$ is calculated 
according to Eqs.~(\ref{eq:Aj}) and (\ref{eq:density}).
The spatial  distributions of $A_j$ along the directions of the dot diameter and dot height are plotted in Figure~\ref{fig:Aj}. 
The maximum value of $A_j$ (7~neV) is found at the As nucleus located at the center of the quantum dot.  
The $A_j$ value associated with an As nucleus is about 1.7 times larger than that associated with the In and Ga nuclei. 
This large difference is due to the larger electron density on anions than on cations. 
The global distribution of $A_j$ reflects the localization of the electron density.  
Although the electron confinement inside the dot is quite effective along the radial direction, 
along the vertical axis the electron density extends farther outside the dot. 
This causes the electron spin to interact with a large number of nuclei outside the dot
in addition to the interaction with the nuclei inside the dot. 
The number of  nuclei for which $A_j$ is larger than $0.01* {\rm max}(A_j)$ is about 60000, 
whereas the number of nuclei inside the dot is about 30000. 

\section{Nuclear magnetic field} \label{sec:BN}

The electron-nuclear spin interaction can be expressed in terms of an effective nuclear magnetic field ${\bf B}_N$ acting on the electron, which is obtained from Eq.~(\ref{eq:HF}) and is defined as 
\ba
{\bf B}_N&=&\frac{1}{g_e \mu_B}  \sum_j A_j  \left\langle {\bf I}_j \right\rangle. \label{eq:BN} 
\ea 
Here $g_e$ is the Land\'{e} factor of the electron and the matrix element $\left\langle {\bf I}_j \right\rangle$ is taken over the nuclear spin state. 
The effective nuclear magnetic field ${\bf B}_N$ is thus determined by the spatial distribution of $A_j$ and by the nuclear spin orientation $\langle {\bf I}_j\rangle$. 
For the quantum dot modeled here, we find that $B_N$ is of the order of 0.01~T when the nuclear spins are unpolarized, and of the order of 1~T when the nuclear spins are polarized. 
The effective nuclear magnetic field ${\bf B}_N$ fluctuates spatially from dot to dot.
The spatial fluctuation arises from inhomogeneities in the environment, such as the random nuclear spin orientation, the dot-size distribution, the alloy and interface disorder. 

When the nuclear spins are unpolarized,  the magnitude and direction of ${\bf B}_N$ are randomly distributed.
The fluctuation of ${\bf B}_N$ due to the random nuclear configuration is given by
\ba
\Delta {\bf B}_N &=&\sqrt{ \langle {\bf B}_N^2\rangle_{\rm ens}  - \langle {\bf B}_N \rangle^2_{\rm ens}} \nonumber \\
                    &=&\frac{1}{g_e\mu_B}\sqrt{\sum A_j^2 (\langle \hat{\bf I}^2_j\rangle_{\rm ens} -\langle \hat{\bf I}_j\rangle^2_{\rm ens})}\nonumber \\
                    &=&\frac{1}{g_e\mu_B}\sqrt{\sum A_j^2I_j(I_j+1)}~ \hat{\bf u}, \label{eq:dBN1}
\ea
where $\langle \cdots \rangle_{\rm ens}$ is an average over the ensemble of dots, and  $\hat{\bf u} $ is the unit vector in a random  direction.
Note that although the nuclear spin ${\bf I}_j$ orientation is changing within the ensemble, the coupling constants $A_j$ remain unchanged under the assumption that dot geometry and atomic configuration are identical for all the dots.  

The fluctuation $\Delta {\bf B}_N$ due to the random distribution of nuclear spin orientations can be suppressed 
by polarizing the nuclear spins. However, even when the nuclear spins are fully polarized, 
the magnitude and direction of ${\bf B}_N$ can still be broadened by other inhomogeneities in the dot array. 
For example, an ensemble of self-assembled quantum dots has typically about a 10\% size distribution, 
which is inherent to the non-equilibrium molecular bean epitaxy growth process.\cite{patella} 
When the dot size changes, the effective number of nuclei interacting with the confined electron and the spatial distribution of $A_j$ change. 
This leads to different values of ${\bf B}_N$ for dots with different sizes.
The fluctuation in ${\bf B}_N$ due to the size distribution will be estimated here by comparing calculated values of 
${\bf B}_N$ for three different dot geometries with base diameter and height values of (14~nm, 5.5~nm),  (15~nm, 6~nm),  and (16~nm, 6.5~nm), respectively. 
From the smallest to the largest dot, the number of nuclei inside the dot increases from 22304 to 35161. 

We further consider two additional sources for the broadening of ${\bf B}_N$ in self-assembled dots: alloy and interface disorder. The alloy disorder stems from the fact that a large number of atomic configurations will yield the same compositional ratio in an InGaAs quantum dot. 
The interfaces between an unalloyed InAs dot and the GaAs buffer shows In-Ga intermixing over a length scale of 1.25~nm, which is the origin of interface disorder.\cite{lita} 
Alloy and interface disorder lead to the broadening of ${\bf B}_N$ in two ways.
First, In and Ga have different nuclear spin quantum numbers ($I_{\rm In}$=4.5, $I_{\rm Ga}$=1.5).
Second, they have different ionic potentials that will lead to a change in the electron densities (or $A_j$).
 
When the nuclear spins are polarized and the dot size is uniform, the fluctuation $\Delta {\bf B}_N$ due to the alloy and interface disorder is given by 
\ba
\Delta {\bf B}_N = \frac{1}{g_e\mu_B}\sqrt{\sum \Delta^2(A_j  I^n_j)  } ~\hat{\bf n},
\ea
where $\hat{\bf n}$ is the unit vector along the nuclear polarization direction,  
$I^n_j$ is the component of ${\bf I}_j$ along $\hat{\bf n}$,
and $\Delta^2(A_j I^n_j)$ is the variance of $A_j I^n_j$.
The variance $\Delta^2(A_j  I^n_j)$  is studied by examining three different atomic configurations for an In$_{1-x}$Ga$_{x}$As dot. 
The three atomic configurations are constructed by randomly choosing the cation atoms as In or Ga with probability $1-x$ and $x$, respectively. 
The overlap between the wave functions for two arbitrary configurations is about 0.997, indicating that 
the fluctuation of the energy density (or $A_j$) is very small. 
Furthermore, the average of the density fluctuation per site is only about $10^{-4}$\% of the average density. 
Therefore, we may justifiably choose to ignore the fluctuation of $A_j$ and approximate $\Delta{\bf B}_N$ as 
\ba
\Delta{\bf B}_N \approx \frac{1}{g_e\mu_B}\sqrt{\sum A_j^2 \Delta^2 I^n_j} ~\hat{\bf n}, \label{eq:dBN2}
\ea 
where $\Delta^2 I^n_j$ is  the variance of $I^n_j$.
For the case of the alloy disorder in  In$_{1-x}$Ga$_x$As dots, $\Delta^2 I^n_j$  is calculated to be $x(1-x)(I_{\rm In}- I_{\rm Ga})^2 $ for all In and Ga atoms and is zero for all As atoms, where $\hat{\bf n}$ is a unit vector along the  nuclear polarization direction. 
For the case of the interface disorder in InAs dots,  $\Delta^2 I^n_j$ is $0.25(I_{\rm In}- I_{\rm Ga})^2$ for the In and Ga atoms in the interface region and is zero for all other atoms. Each cation atom site in the interface is taken to have probability 0.5 to be occupied by either an In or a Ga atom. 

After calculating the fluctuation of ${\bf B}_N$ due to the inhomogeneities in the environment as described above, we have obtained the following results. 
When the nuclear spins are unpolarized, a random nuclear spin configuration yields a value for $\Delta {\bf B}_N$ 
on the order of 100~G.  
When the nuclear spins are polarized, an 10\% dot-size distribution also yields  $\Delta {\bf B}_N$ of the order of 100~G. 
When the nuclear spins are polarized and the dot size is uniform, the alloy disorder results in fluctuations on the order of 10~G, while the interface disorder gives rise to fluctuations of the order of 0.1~G.
These results indicate that an unpolarized nuclear spin configuration and quantum dot size fluctuation are the dominant sources of the inhomogeneous nuclear magnetic field. 
Since the electron localized in each quantum dot is immersed in a different magnetic field ${\bf B}_N$, 
the Zeeman splitting ($E=g_e\mu_B {\bf B}_N \cdot {\bf S}$) and precession frequency $g_e\mu_B B_N/\hbar$ of each electron spin are different. 
This leads to a fluctuation in the Zeeman splitting ($\Delta E$) and an ensemble dephasing with  
a dephasing time $T_2^*$ defined as $\hbar/g_e\mu_B \Delta B_N$. 
The  values of $\Delta {\bf B}_N$, $\Delta E$, and $T_2^*$ resulting from the inhomogeneous environments studied here are summarized in Table~\ref{tab:dephasing}. 

\begin{table}[t]
\caption{Nuclear magnetic field spatial fluctuation ($\Delta B_N$),  Zeeman energy fluctuation
$(\Delta E= g_e\mu_B \Delta B_N)$,  and dephasing time ($T_2^*=\hbar/g_e\mu_B\Delta B_N$) 
caused by various inhomogeneities for an InAs quantum dot embedded in a GaAs buffer. 
For unpolarized nuclei, each nuclear spin direction is chosen randomly. 
For dot-size fluctuations,  the  base diameter is set to 15$\pm$1~nm and the height to 6$\pm$0.5~nm. 
For alloy disorder, In$_{0.5}$Ga$_{0.5}$As dots are examined and each cation atom is randomly chosen to be an In or a Ga atom. 
For interface disorder, each cation within a 1.25~nm thick interface between the dot and the buffer is randomly chosen to be an In or a Ga atom, reflecting the experimental observation of In-Ga mixing near the interface (Ref.~\onlinecite{lita}).} 
\label{tab:dephasing}
\begin{ruledtabular}
\begin{tabular}{cccc}
Inhomogeneous Environment & $\Delta B_N$ (G)  &  $\Delta E$ (eV) & $T_2^*$ (s) \\
\hline
Unpolarized nuclei & 100 &  $10^{-6}$ & $10^{-10}$ \\
Dot-size fluctuation  & 100 &  $10^{-6}$ &  $10^{-10}$ \\
Alloy disorder & 10 &  $10^{-7}$  &  $10^{-9}$ \\
Interface disorder & 0.1& $10^{-9}$  &   $10^{-7}$
\end{tabular}
\end{ruledtabular}
\end{table}

\section{Effect of nuclear magnetic field on qubit operations} \label{sec:operation}

First, we examine the effect of ${\bf B}_N$ on a two-qubit operation.
The fluctuation of ${\bf B}_N$ leads to the inequality of the Zeeman energy in the two qubits. When a two-qubit operation such as a ``swap'' operation uses the exchange interaction $J {\bf S}_1 \cdot {\bf S}_2$, the Zeeman-energy difference $\Delta E_Z$ between two qubits causes an error $\sim(\Delta E_Z/J)^2$.\cite{hu-sousa-sarma}   
For error correction codes to be effective,\cite{steane}  we require $(\Delta E/ J)^2 < 10^{-4}$. 
The Zeeman-energy fluctuation $\Delta E_Z$ due to the four different sources of inhomogeneity in $B_N$ considered in this paper ranges from 1~neV to 1~$\mu$eV. 
Hence,  the exchange energy $J$ should be larger than 100~$\mu$eV.
At the same time, to prevent the electron from being excited to higher-lying orbitals, 
$J$ should be smaller than the electron energy spacing ($\Delta E_e$) between the ground and the excited orbital.  
The excitation probability due to the exchange interaction is roughly on the order of $(J/\Delta E_e)^2$. 
Therefore, $J/\Delta E_e < 10^{-2}$ would ensure the electron to stay in the qubit space with leakage probability below $10^{-4}$. 

The dual condition ($\Delta E_z \ll J \ll \Delta E_e$) can be met with vertically-stacked self-assembled dots.\cite{ibanez}
A recent calculation with harmonic double-well confinement potentials suggests that $J$ can be varied from 10~meV to 0.1~meV 
as the inter-dot distance increases from 5~nm to  20~nm.\cite{burkard} 
Self-assembled dots with vertical inter-dot distance as small as 2~nm can be easily fabricated.\cite{ibanez}
With a given physical inter-dot distance, an effective inter-dot distance can be electronically tuned with gate voltages 
to turn on and off the exchange interaction.  
The electron energy spacing $\Delta E_e$ of a self-assembled dot is 
about 50 -- 100~meV, depending on the geometry and size of the dot.\cite{heitz} 
Therefore, $J$ that satisfies the dual condition is between 0.1--1~meV, which is achievable with 
vertically-stacked self-assembled dots. In conclusion, with $J$ between 0.1 -- 1~meV, 
the error due to the inhomogeneous Zeeman energies is smaller than the threshold for error correction, 
and the qubit leakage to higher orbitals is effectively prevented.     

Second, we consider the effect of ${\bf B}_N$ on a single-qubit operation.
The single-qubit operation using the Zeeman coupling to an electron spin resonance (ESR) 
field ($B_{ac} \cos{\omega_{ac}t}$) involves the tuning of the ESR field frequency to the electron-spin precession frequency or vice versa. 
This tuning can be achieved by applying a gate voltage in order to make the electron wave function overlap with a material having a different g factor.\cite{loss-divincenzo, vrijen} The tuning process becomes complicated in the presence of ${\bf B}_N$, which affects the precession frequency of the electron spin. 
The effective nuclear magnetic field ${\bf B}_N$ fluctuates in space from one qubit to another and evolves in time. 
The spatial fluctuation of  $B_N$ can be compensated by calibrating the gate voltage for each qubit separately.
However, the temporal evolution of $B_N$ is difficult to compensate since a gate calibration cannot be done 
immediately before each operation. 
The temporal evolution is determined by 
many competing interactions such as the Zeeman coupling of the nuclear spin to 
the external magnetic field, the nuclear spin interaction 
with the electron spin, the nuclear spin dipolar interaction, and the nuclear spin-lattice interaction. 
Detailed studies of the temporal evolution of ${\bf B}_N$ that include these interactions are needed to determine 
how long a single-qubit gate calibration is valid. 

Here, we estimate the upper limit of the temporal change of ${\bf B}_N$ 
for a single-qubit gate calibration to be valid. We assume that a static magnetic field $B_0$ of the order of 1~T is applied, 
and that an  ESR field of the order of 0.001~T is used for the spin rotation.\cite{kane,vrijen,loss-divincenzo}
The precession frequency of the electron spin is given by $\omega_e = g_e \mu_B \sqrt { (B_0+B_N^{||})^2 + (B_N^{\perp})^ 2}$, where $B_N^{||}$ and $B_N^{\perp}$ are the ${\bf B}_N$ component parallel and perpendicular to $B_0$, respectively. A single-qubit gate will be calibrated by tuning the frequency of the ESR field $\omega_{ac}$ to $\omega_e$. After some time,  $B_N^{||}$ and $B_N^{\perp}$ will change by $\Delta B_N^{||}$ and $\Delta B_N^{\perp}$ due to nuclear spin dynamics, and  $\omega_e$ will change accordingly. This will lead to the detuning of the ESR field. 
The frequency difference $\omega_{ac}-\omega_e$  is approximately $g_e\mu_B (\Delta B_N^{||} +
(B_N^{\perp}/B_0) \Delta B_N^{\perp})$,  using that $B_0 \gg B_N$ for unpolarized nuclear spins where $B_N$ is on the order of 0.01~T.
The error due to the detuning in the single-qubit operation is proportional to $(\omega_{ac}-\omega_e)^2/(g_e\mu_B B_{ac})^2$. To have the error to be smaller than the error threshold ($10^{-4}$) given by the error correction code,\cite{steane}  
$\Delta B_N^{||}$ and $\Delta B_N^{\perp}$ should be smaller than $10^{-5}$~T and $10^{-3}$~T, respectively.
If  $\Delta B_N^{||}$ and $\Delta B_N^{\perp}$ during the time interval 
between the calibration and the end of QC are bigger than these upper limits, 
the single-qubit gate calibration becomes invalid, and thus 
QC architectures using only exchange interactions should be employed.\cite{kempe1,kempe2, divincenzo-bacon} 
The exchange interaction alone can provide universal quantum gates with a minimal cost of increasing the number of required 
physical qubits by three times and of increasing the number of gate operations by five to seven times.\cite{kempe2} 
          
\section{Conclusion} \label{sec:conclusion}
   
The effect of electron-nuclear spin interactions on qubit operations is investigated 
here for a qubit represented by the electron spin localized in a self-assembled quantum dot.
The localized electron wave function is evaluated within the atomistic tight-binding model, 
and the magnetic field generated by hyperfine coupling to the nuclear spins is estimated in the presence of 
an inhomogeneous environment characterized by random nuclear configurations, 
dot-size fluctuations, and alloy and interface disorder. 
Due to each of these inhomogeneities, the effective magnetic field felt by each electron varies on the order of 100~G, 100~G, 10~G, and 0.1~G, respectively. 

The inhomogeneous nuclear magnetic field causes  an error in two-qubit operations due 
to the inequality of the Zeeman splitting in the two qubits. 
However, the errors can be made smaller than the quantum error threshold, 
as long as the exchange energy for the two-qubit operation is larger than 0.1~meV.  
Recent work indicates that an exchange energy of 0.1~meV or larger is easily achievable with vertically stacked quantum dots.\cite{son, bayer, burkard} 
At the same time, the large energy spacing between the ground and the excited orbital (50--100~meV)
of the quantum dots 
ensures that the electron qubit stays in the ground orbital while the two-qubit operation is conducted.\cite{heitz} 
We also present the upper limit of the temporal change of the nuclear magnetic field for a single-qubit gate calibration to be valid
when an electron resonance field is used for a single-qubit operation. 
The changes of the nuclear magnetic field parallel and perpendicular to the external static magnetic field should be smaller than 
$10^{-5}$~T and $10^{-3}$~T, respectively. 
When this condition is not met, QC architectures using only exchange interactions 
should be employed.\cite{kempe1,kempe2,divincenzo-bacon} 
Using the exchange interaction which can provide universal quantum gates, 
quantum-dot based quantum computer architectures are 
scalable to many qubits even in the presence of inhomogeneous environments causing 
fluctuations in the effective nuclear magnetic field. 

\begin{acknowledgements} This work was performed at Jet Propulsion
Laboratory, California Institute of Technology under a contract with the
National Aeronautics and Space Administration. This work was supported by
grants from NSA/ARDA, ONR, and JPL internal Research and Development.  
\end{acknowledgements}

\appendix*

\section{Density of an atomic basis orbital at a nuclear site}
In principle, a measurement of the Overhauser shift of the electron spin resonance provides the density of the conduction electron at the nuclear site. 
Although Overhauser shifts for bulk InAs and GaAs have not been measured, 
the densities of an atomic orbital at a nuclear site for InAs and GaAs can be deduced from the Overhauser shifts measured for bulk InSb.\cite{gueron} The measured densities for InSb are
$|\psi(0)|^2_{(\rm In~in~InSb)} = 9.4 \times 10^{25}$ cm$^{-3}$ and 
$|\psi(0)|^2_{(\rm Sb~in~InSb)} = 1.6\times 10^{26}$ cm$^{-3}$. 
Bulk InAs and GaAs have an ionicity similar to that of InSb (0.36, 0.31, and 0.32, respectively).\cite{phillips} 
Therefore, the distribution of the valence electrons between the anion and cation atoms is similar 
in all III-V semiconductor materials.
This implies that the ratio of the valence electron densities in bulk is the same as the ratio of the densities for individual atoms:\cite{paget}.
Finally, as a first-order approximation we also assume that this is true for the ratio of the conduction electron densities:
\ba
\frac{|\psi(0)|^2_{(\rm In~in~InAs)}}{  |\psi(0)|^2_{(\rm In~in~InSb)} } = 
\frac{|\psi(0)|^2_{(\rm In~ atom)}}{  |\psi(0)|^2_{(\rm In~ atom)} }, \\
\frac{|\psi(0)|^2_{(\rm As~in~InAs)}}{  |\psi(0)|^2_{(\rm Sb~in~InSb)} } = 
\frac{|\psi(0)|^2_{(\rm As ~atom)}}{  |\psi(0)|^2_{(\rm Sb ~atom)} }, \\
\frac{|\psi(0)|^2_{(\rm Ga~in~GaAs)}}{ |\psi(0)|^2_{(\rm In~in~InSb)} } 
= \frac{|\psi(0)|^2_{(\rm Ga~ atom)}}{ |\psi(0)|^2_{(\rm In ~atom)} }, \\
\frac{|\psi(0)|^2_{(\rm As~in~GaAs)}}{ |\psi(0)|^2_{(\rm Sb~in~InSb)} } 
= \frac{|\psi(0)|^2_{(\rm As ~atom)}}{ |\psi(0)|^2_{(\rm Sb~ atom)} }.
\ea
Since the atomic density ratios are known,\cite{bennett} the conduction electron densities 
in bulk InAs and GaAs can then be deduced:
\ba
|\psi(0)|^2_{(\rm In~in~InAs)} \simeq 9.4 \times 10^{25} {\rm cm}^{-3}, \\  
|\psi(0)|^2_{(\rm As~in~InAs)} \simeq 9.8 \times 10^{25} {\rm cm}^{-3}, \\ 
|\psi(0)|^2_{(\rm Ga~in~GaAs)} \simeq 5.8 \times 10^{25} {\rm cm}^{-3}, \\ 
|\psi(0)|^2_{(\rm As~in~GaAs)} \simeq 9.8 \times 10^{25} {\rm cm}^{-3}.
\ea

The density $|\psi(0)|^2$ for each atom in bulk is  related to the tight-binding orbitals $\phi_{s^*}(0)$ and $\phi_{s}(0)$ as follows:
\ba
|\psi(0)|^2 = |\alpha \phi_{s}(0) + \beta \phi_{s^*}(0) |^2, \label{eq:appendix}
\ea
where $\alpha$ and $\beta$ are the tight-binding coefficients for the conduction-band edge wave function in bulk.
These coefficients are listed in Table~\ref{tab:appendix}.
Determining the unknown values $\phi_s(0)$ and $\phi_{s^*}(0)$ requires one more equation in addition to Eq.~(\ref{eq:appendix}). 
We assume that the ratio $\phi_{s^*}(0)/\phi_{s}(0)$ is equal to the ratio of the corresponding atomic orbitals. 
The atomic orbital can be described by a hydrogen-like atomic orbital with
an effective nuclear charge.\cite{clementi-raimondi, clementi-raimondi-reinhardt}
The ratios $\phi_{s^*}(0)/\phi_{s}(0)$ for In, Ga, and As atoms are 0.53, 0.44, and 0.30, respectively.
Finally, by inserting the deduced density $|\psi(0)|^2$, the conduction-band edge coefficients $\alpha, \beta$,
and the orbital ratio $\phi_{s^*}(0)/\phi_{s}(0)$ into Eq.~(\ref{eq:appendix}),  
the densities of the $s$ and $s^*$ orbitals are obtained. The resulting densities are listed in Table~\ref{tab:appendix}.
 
\begin{table}[t]
\caption{Tight-binding coefficients of the $s$ and $s^*$ basis orbitals for the conduction-band-edge
wave functions of bulk InAs and GaAs, and deduced densities of $s$ and $s^*$ orbitals at nuclear sites. 
The densities $|\phi_s(0)|^2$ and $|\phi_{s^*}(0)|^2$ are in units of $10^{25} {\rm cm}^{-3}$.} 
\label{tab:appendix}
\begin{ruledtabular}
\begin{tabular}{ccccc}
Atom & $\alpha$ &  $\beta$ & $|\phi_s(0)|^2$  & $|\phi_{s^*}(0)|^2$  \\
\hline
In in InAs & 0.974 & 0.228 & 7.9 &  2.2 \\
As in InAs & 0.872 & -0.489 & 18.4 & 1.7 \\
Ga in GaAs & 0.988 & 0.157 & 5.2 & 1.0 \\
As in GaAs & 0.869 & -0.576 & 20.2 & 1.8 
\end{tabular}
\end{ruledtabular}
\end{table}


\begin{thebibliography}{37}
\expandafter\ifx\csname natexlab\endcsname\relax\def\natexlab#1{#1}\fi
\expandafter\ifx\csname bibnamefont\endcsname\relax
  \def\bibnamefont#1{#1}\fi
\expandafter\ifx\csname bibfnamefont\endcsname\relax
  \def\bibfnamefont#1{#1}\fi
\expandafter\ifx\csname citenamefont\endcsname\relax
  \def\citenamefont#1{#1}\fi
\expandafter\ifx\csname url\endcsname\relax
  \def\url#1{\texttt{#1}}\fi
\expandafter\ifx\csname urlprefix\endcsname\relax\def\urlprefix{URL }\fi
\providecommand{\bibinfo}[2]{#2}
\providecommand{\eprint}[2][]{\url{#2}}

\bibitem[{\citenamefont{Gershenfeld and Chuang}(1997)}]{gershenfeld}
\bibinfo{author}{\bibfnamefont{N.~A.} \bibnamefont{Gershenfeld}}
  \bibnamefont{and} \bibinfo{author}{\bibfnamefont{I.~L.}
  \bibnamefont{Chuang}}, \bibinfo{journal}{Science}
  \textbf{\bibinfo{volume}{275}}, \bibinfo{pages}{350} (\bibinfo{year}{1997}).

\bibitem[{\citenamefont{Yamaguchi and Yamamoto}(1999)}]{yamaguchi}
\bibinfo{author}{\bibfnamefont{F.}~\bibnamefont{Yamaguchi}} \bibnamefont{and}
  \bibinfo{author}{\bibfnamefont{Y.}~\bibnamefont{Yamamoto}},
  \bibinfo{journal}{Appl. Phys. A} \textbf{\bibinfo{volume}{68}},
  \bibinfo{pages}{1} (\bibinfo{year}{1999}).

\bibitem[{\citenamefont{Kane}(1998)}]{kane}
\bibinfo{author}{\bibfnamefont{B.~E.} \bibnamefont{Kane}},
  \bibinfo{journal}{Nature} \textbf{\bibinfo{volume}{393}},
  \bibinfo{pages}{133} (\bibinfo{year}{1998}).

\bibitem[{\citenamefont{Twamley}(2003)}]{twamley}
\bibinfo{author}{\bibfnamefont{J.}~\bibnamefont{Twamley}},
  \bibinfo{journal}{Phys. Rev. A} \textbf{\bibinfo{volume}{67}},
  \bibinfo{pages}{052318} (\bibinfo{year}{2003}).

\bibitem[{\citenamefont{Loss and DiVincenzo}(1998)}]{loss-divincenzo}
\bibinfo{author}{\bibfnamefont{D.}~\bibnamefont{Loss}} \bibnamefont{and}
  \bibinfo{author}{\bibfnamefont{D.~P.} \bibnamefont{DiVincenzo}},
  \bibinfo{journal}{Phys. Rev. A} \textbf{\bibinfo{volume}{57}},
  \bibinfo{pages}{120} (\bibinfo{year}{1998}).

\bibitem[{\citenamefont{Imamoglu et~al.}(1999)\citenamefont{Imamoglu,
  Awschalom, Burkard, DiVincenzo, Loss, Sherwin, and Small}}]{imamoglu}
\bibinfo{author}{\bibfnamefont{A.}~\bibnamefont{Imamoglu}},
  \bibinfo{author}{\bibfnamefont{D.~D.} \bibnamefont{Awschalom}},
  \bibinfo{author}{\bibfnamefont{G.}~\bibnamefont{Burkard}},
  \bibinfo{author}{\bibfnamefont{D.~P.} \bibnamefont{DiVincenzo}},
  \bibinfo{author}{\bibfnamefont{D.}~\bibnamefont{Loss}},
  \bibinfo{author}{\bibfnamefont{M.}~\bibnamefont{Sherwin}}, \bibnamefont{and}
  \bibinfo{author}{\bibfnamefont{A.}~\bibnamefont{Small}},
  \bibinfo{journal}{Phys. Rev. Lett.} \textbf{\bibinfo{volume}{83}},
  \bibinfo{pages}{4204} (\bibinfo{year}{1999}).

\bibitem[{\citenamefont{Vrijen et~al.}(2000)\citenamefont{Vrijen, Yablonovitch,
  Wang, Jiang, Balandin, Roychowdhury, Mor, and DiVincenzo}}]{vrijen}
\bibinfo{author}{\bibfnamefont{R.}~\bibnamefont{Vrijen}},
  \bibinfo{author}{\bibfnamefont{E.}~\bibnamefont{Yablonovitch}},
  \bibinfo{author}{\bibfnamefont{K.}~\bibnamefont{Wang}},
  \bibinfo{author}{\bibfnamefont{H.~W.}~\bibnamefont{Jiang}},
  \bibinfo{author}{\bibfnamefont{A.}~\bibnamefont{Balandin}},
  \bibinfo{author}{\bibfnamefont{V.}~\bibnamefont{Roychowdhury}},
  \bibinfo{author}{\bibfnamefont{T.}~\bibnamefont{Mor}}, \bibnamefont{and}
  \bibinfo{author}{\bibfnamefont{D.}~\bibnamefont{DiVincenzo}},
  \bibinfo{journal}{Phys. Rev. A} \textbf{\bibinfo{volume}{62}},
  \bibinfo{pages}{012306} (\bibinfo{year}{2000}).

\bibitem[{\citenamefont{Chuang et~al.}(1998)\citenamefont{Chuang, Vandersypen,
  Zhou, Leung, and Lloyd}}]{chuang-vandersypen}
\bibinfo{author}{\bibfnamefont{I.~L.} \bibnamefont{Chuang}},
  \bibinfo{author}{\bibfnamefont{L.~M.~K.} \bibnamefont{Vandersypen}},
  \bibinfo{author}{\bibfnamefont{X.}~\bibnamefont{Zhou}},
  \bibinfo{author}{\bibfnamefont{D.~W.} \bibnamefont{Leung}}, \bibnamefont{and}
  \bibinfo{author}{\bibfnamefont{S.}~\bibnamefont{Lloyd}},
  \bibinfo{journal}{Nature} \textbf{\bibinfo{volume}{393}},
  \bibinfo{pages}{143} (\bibinfo{year}{1998}).

\bibitem[{\citenamefont{Vandersypen et~al.}(2000)\citenamefont{Vandersypen,
  Steffen, Breyta, Yannoni, Cleve, and Chuang}}]{vandersypen-steffen-prl}
\bibinfo{author}{\bibfnamefont{L.~M.~K.} \bibnamefont{Vandersypen}},
  \bibinfo{author}{\bibfnamefont{M.}~\bibnamefont{Steffen}},
  \bibinfo{author}{\bibfnamefont{G.}~\bibnamefont{Breyta}},
  \bibinfo{author}{\bibfnamefont{C.~S.} \bibnamefont{Yannoni}},
  \bibinfo{author}{\bibfnamefont{R.}~\bibnamefont{Cleve}}, \bibnamefont{and}
  \bibinfo{author}{\bibfnamefont{I.~L.} \bibnamefont{Chuang}},
  \bibinfo{journal}{Phys. Rev. Lett.} \textbf{\bibinfo{volume}{85}},
  \bibinfo{pages}{5452} (\bibinfo{year}{2000}).

\bibitem[{\citenamefont{Vandersypen et~al.}(2001)\citenamefont{Vandersypen,
  Steffen, Breyta, Yannoni, Sherwood, and Chuang}}]{vandersypen-steffen-nature}
\bibinfo{author}{\bibfnamefont{L.~M.~K.} \bibnamefont{Vandersypen}},
  \bibinfo{author}{\bibfnamefont{M.}~\bibnamefont{Steffen}},
  \bibinfo{author}{\bibfnamefont{G.}~\bibnamefont{Breyta}},
  \bibinfo{author}{\bibfnamefont{C.~S.} \bibnamefont{Yannoni}},
  \bibinfo{author}{\bibfnamefont{M.~H.} \bibnamefont{Sherwood}},
  \bibnamefont{and} \bibinfo{author}{\bibfnamefont{I.~L.}
  \bibnamefont{Chuang}}, \bibinfo{journal}{Nature}
  \textbf{\bibinfo{volume}{414}}, \bibinfo{pages}{883} (\bibinfo{year}{2001}).

\bibitem[{\citenamefont{Gulde et~al.}(2003)\citenamefont{Gulde, Riebe,
  Lancaster, Becher, Eschner, Haffner, Schmidt-Kaler, Chuang, and
  Blatt}}]{gulde}
\bibinfo{author}{\bibfnamefont{S.}~\bibnamefont{Gulde}},
  \bibinfo{author}{\bibfnamefont{M.}~\bibnamefont{Riebe}},
  \bibinfo{author}{\bibfnamefont{G.~P.~T.} \bibnamefont{Lancaster}},
  \bibinfo{author}{\bibfnamefont{C.}~\bibnamefont{Becher}},
  \bibinfo{author}{\bibfnamefont{J.}~\bibnamefont{Eschner}},
  \bibinfo{author}{\bibfnamefont{H.}~\bibnamefont{Haffner}},
  \bibinfo{author}{\bibfnamefont{F.}~\bibnamefont{Schmidt-Kaler}},
  \bibinfo{author}{\bibfnamefont{I.~L.} \bibnamefont{Chuang}},
  \bibnamefont{and} \bibinfo{author}{\bibfnamefont{R.}~\bibnamefont{Blatt}},
  \bibinfo{journal}{Nature} \textbf{\bibinfo{volume}{421}}, \bibinfo{pages}{48}
  (\bibinfo{year}{2003}).

\bibitem[{\citenamefont{Steane}(2003)}]{steane}
\bibinfo{author}{\bibfnamefont{A.~M.} \bibnamefont{Steane}},
  \bibinfo{journal}{Phys. Rev. A} \textbf{\bibinfo{volume}{68}},
  \bibinfo{pages}{042322} (\bibinfo{year}{2003}).

\bibitem[{\citenamefont{Hu et~al.}(2001)\citenamefont{Hu, de~Sousa, and
  Sarma}}]{hu-sousa-sarma}
\bibinfo{author}{\bibfnamefont{X.}~\bibnamefont{Hu}},
  \bibinfo{author}{\bibfnamefont{R.}~\bibnamefont{de~Sousa}}, \bibnamefont{and}
  \bibinfo{author}{\bibfnamefont{S.}~\bibnamefont{Das~Sarma}},
  \bibinfo{journal}{Phys. Rev. Lett.} \textbf{\bibinfo{volume}{86}},
  \bibinfo{pages}{918} (\bibinfo{year}{2001}).

\bibitem[{\citenamefont{Boykin et~al.}(2002)\citenamefont{Boykin, Klimeck,
  Bowen, and Oyafuso}}]{boykin_strain}
\bibinfo{author}{\bibfnamefont{T.~B.} \bibnamefont{Boykin}},
  \bibinfo{author}{\bibfnamefont{G.}~\bibnamefont{Klimeck}},
  \bibinfo{author}{\bibfnamefont{R.~C.} \bibnamefont{Bowen}}, \bibnamefont{and}
  \bibinfo{author}{\bibfnamefont{F.}~\bibnamefont{Oyafuso}},
  \bibinfo{journal}{Phys. Rev. B} \textbf{\bibinfo{volume}{66}},
  \bibinfo{pages}{125207} (\bibinfo{year}{2002}).

\bibitem[{\citenamefont{Moison et~al.}(1994)\citenamefont{Moison, Houzay,
  Barthe, Leprince, Andr\'e, and Vatel}}]{moison}
\bibinfo{author}{\bibfnamefont{J.~M.} \bibnamefont{Moison}},
  \bibinfo{author}{\bibfnamefont{F.}~\bibnamefont{Houzay}},
  \bibinfo{author}{\bibfnamefont{F.}~\bibnamefont{Barthe}},
  \bibinfo{author}{\bibfnamefont{L.}~\bibnamefont{Leprince}},
  \bibinfo{author}{\bibfnamefont{E.}~\bibnamefont{Andr\'e}}, \bibnamefont{and}
  \bibinfo{author}{\bibfnamefont{O.}~\bibnamefont{Vatel}},
  \bibinfo{journal}{Appl. Phys. Lett.} \textbf{\bibinfo{volume}{64}},
  \bibinfo{pages}{196} (\bibinfo{year}{1994}).

\bibitem[{\citenamefont{Kobayashi et~al.}(1996)\citenamefont{Kobayashi,
  Ramachandran, Chen, and Madhukar}}]{kobayashi}
\bibinfo{author}{\bibfnamefont{N.~P.} \bibnamefont{Kobayashi}},
  \bibinfo{author}{\bibfnamefont{T.~R.} \bibnamefont{Ramachandran}},
  \bibinfo{author}{\bibfnamefont{P.}~\bibnamefont{Chen}}, \bibnamefont{and}
  \bibinfo{author}{\bibfnamefont{A.}~\bibnamefont{Madhukar}},
  \bibinfo{journal}{Appl. Phys. Lett.} \textbf{\bibinfo{volume}{68}},
  \bibinfo{pages}{3299} (\bibinfo{year}{1996}).

\bibitem[{\citenamefont{Mukahametzhanov
  et~al.}(1999)\citenamefont{Mukahametzhanov, Wei, Heitz, and
  Madhukar}}]{mukhametzhanov}
\bibinfo{author}{\bibfnamefont{I.}~\bibnamefont{Mukahametzhanov}},
  \bibinfo{author}{\bibfnamefont{J.}~\bibnamefont{Wei}},
  \bibinfo{author}{\bibfnamefont{R.}~\bibnamefont{Heitz}}, \bibnamefont{and}
  \bibinfo{author}{\bibfnamefont{A.}~\bibnamefont{Madhukar}},
  \bibinfo{journal}{Appl. Phys. Lett.} \textbf{\bibinfo{volume}{75}},
  \bibinfo{pages}{85} (\bibinfo{year}{1999}).

\bibitem[{\citenamefont{Keating}(1966)}]{keating}
\bibinfo{author}{\bibfnamefont{P.}~\bibnamefont{Keating}},
  \bibinfo{journal}{Phys. Rev.} \textbf{\bibinfo{volume}{145}},
  \bibinfo{pages}{637} (\bibinfo{year}{1966}).

\bibitem[{\citenamefont{L\"owdin}(1950)}]{lowdin}
\bibinfo{author}{\bibfnamefont{P.-O.} \bibnamefont{L\"owdin}},
  \bibinfo{journal}{J. Chem. Phys.} \textbf{\bibinfo{volume}{18}},
  \bibinfo{pages}{365} (\bibinfo{year}{1950}).

\bibitem[{\citenamefont{Harrison}(1999)}]{harrison}
\bibinfo{author}{\bibfnamefont{W.~A.} \bibnamefont{Harrison}},
  \emph{\bibinfo{title}{Elementary Electronic Structure}}
  (\bibinfo{publisher}{World Scientific, New Jersey}, \bibinfo{year}{1999}).

\bibitem[{\citenamefont{Slater and Koster}(1954)}]{slater-koster}
\bibinfo{author}{\bibfnamefont{J.~C.} \bibnamefont{Slater}} \bibnamefont{and}
  \bibinfo{author}{\bibfnamefont{G.~F.} \bibnamefont{Koster}},
  \bibinfo{journal}{Phys. Rev.} \textbf{\bibinfo{volume}{94}},
  \bibinfo{pages}{1498} (\bibinfo{year}{1954}).

\bibitem[{\citenamefont{Gueron}(1964)}]{gueron}
\bibinfo{author}{\bibfnamefont{M.}~\bibnamefont{Gueron}},
  \bibinfo{journal}{Phys. Rev.} \textbf{\bibinfo{volume}{135}},
  \bibinfo{pages}{200} (\bibinfo{year}{1964}).

\bibitem[{\citenamefont{Patella et~al.}(2001)\citenamefont{Patella, Fanfoni,
  Arciprete, Nufris, Placidi, and Balzarotti}}]{patella}
\bibinfo{author}{\bibfnamefont{F.}~\bibnamefont{Patella}},
  \bibinfo{author}{\bibfnamefont{M.}~\bibnamefont{Fanfoni}},
  \bibinfo{author}{\bibfnamefont{F.}~\bibnamefont{Arciprete}},
  \bibinfo{author}{\bibfnamefont{S.}~\bibnamefont{Nufris}},
  \bibinfo{author}{\bibfnamefont{E.}~\bibnamefont{Placidi}}, \bibnamefont{and}
  \bibinfo{author}{\bibfnamefont{A.}~\bibnamefont{Balzarotti}},
  \bibinfo{journal}{Appl. Phys. Lett.} \textbf{\bibinfo{volume}{78}},
  \bibinfo{pages}{320} (\bibinfo{year}{2001}).

\bibitem[{\citenamefont{Lita et~al.}(1999)\citenamefont{Lita, Goldman,
  Phillips, and Bhattacharya}}]{lita}
\bibinfo{author}{\bibfnamefont{B.}~\bibnamefont{Lita}},
  \bibinfo{author}{\bibfnamefont{R.~S.} \bibnamefont{Goldman}},
  \bibinfo{author}{\bibfnamefont{J.~D.} \bibnamefont{Phillips}},
  \bibnamefont{and} \bibinfo{author}{\bibfnamefont{P.~K.}
  \bibnamefont{Bhattacharya}}, \bibinfo{journal}{Appl. Phys. Lett.}
  \textbf{\bibinfo{volume}{75}}, \bibinfo{pages}{2797} (\bibinfo{year}{1999}).

\bibitem[{\citenamefont{Ibanez et~al.}(2003)\citenamefont{Ibanez, Patane,
  Henini, Eaves, Hernandez, Cusco, Artus, Musikhin, and Brounkov}}]{ibanez}
\bibinfo{author}{\bibfnamefont{J.}~\bibnamefont{Ibanez}},
  \bibinfo{author}{\bibfnamefont{A.}~\bibnamefont{Patane}},
  \bibinfo{author}{\bibfnamefont{M.}~\bibnamefont{Henini}},
  \bibinfo{author}{\bibfnamefont{L.}~\bibnamefont{Eaves}},
  \bibinfo{author}{\bibfnamefont{S.}~\bibnamefont{Hernandez}},
  \bibinfo{author}{\bibfnamefont{R.}~\bibnamefont{Cusco}},
  \bibinfo{author}{\bibfnamefont{L.}~\bibnamefont{Artus}},
  \bibinfo{author}{\bibfnamefont{Y.~G.} \bibnamefont{Musikhin}},
  \bibnamefont{and} \bibinfo{author}{\bibfnamefont{P.~N.}
  \bibnamefont{Brounkov}}, \bibinfo{journal}{Appl. Phys. Lett.}
  \textbf{\bibinfo{volume}{83}}, \bibinfo{pages}{3069} (\bibinfo{year}{2003}).

\bibitem[{\citenamefont{Burkard et~al.}(1999)\citenamefont{Burkard, Loss, and
  DiVincenzo}}]{burkard}
\bibinfo{author}{\bibfnamefont{G.}~\bibnamefont{Burkard}},
  \bibinfo{author}{\bibfnamefont{D.}~\bibnamefont{Loss}}, \bibnamefont{and}
  \bibinfo{author}{\bibfnamefont{D.~P.} \bibnamefont{DiVincenzo}},
  \bibinfo{journal}{Phys. Rev. B} \textbf{\bibinfo{volume}{59}},
  \bibinfo{pages}{2070} (\bibinfo{year}{1999}).

\bibitem[{\citenamefont{Heitz et~al.}(2000)\citenamefont{Heitz, Stier,
  Mukhametzhanov, Madhukar, and Bimberg}}]{heitz}
\bibinfo{author}{\bibfnamefont{R.}~\bibnamefont{Heitz}},
  \bibinfo{author}{\bibfnamefont{O.}~\bibnamefont{Stier}},
  \bibinfo{author}{\bibfnamefont{I.}~\bibnamefont{Mukhametzhanov}},
  \bibinfo{author}{\bibfnamefont{A.}~\bibnamefont{Madhukar}}, \bibnamefont{and}
  \bibinfo{author}{\bibfnamefont{D.}~\bibnamefont{Bimberg}},
  \bibinfo{journal}{Phys. Rev. B} \textbf{\bibinfo{volume}{62}},
  \bibinfo{pages}{11017} (\bibinfo{year}{2000}).

\bibitem[{\citenamefont{Kempe et~al.}(2001)\citenamefont{Kempe, Bacon, Lidar,
  and Whaley}}]{kempe1}
\bibinfo{author}{\bibfnamefont{J.}~\bibnamefont{Kempe}},
  \bibinfo{author}{\bibfnamefont{D.}~\bibnamefont{Bacon}},
  \bibinfo{author}{\bibfnamefont{D.~A.} \bibnamefont{Lidar}}, \bibnamefont{and}
  \bibinfo{author}{\bibfnamefont{K.~B.} \bibnamefont{Whaley}},
  \bibinfo{journal}{Phys. Rev. A} \textbf{\bibinfo{volume}{63}},
  \bibinfo{pages}{042307} (\bibinfo{year}{2001}).

\bibitem[{\citenamefont{Kempe and Whaley}(2002)}]{kempe2}
\bibinfo{author}{\bibfnamefont{J.}~\bibnamefont{Kempe}} \bibnamefont{and}
  \bibinfo{author}{\bibfnamefont{K.~B.} \bibnamefont{Whaley}},
  \bibinfo{journal}{Phys. Rev. A} \textbf{\bibinfo{volume}{65}},
  \bibinfo{pages}{052330} (\bibinfo{year}{2002}).

\bibitem[{\citenamefont{DiVincenzo et~al.}(2000)\citenamefont{DiVincenzo,
  Bacon, Kempe, Burkard, and Whaley}}]{divincenzo-bacon}
\bibinfo{author}{\bibfnamefont{D.~P.} \bibnamefont{DiVincenzo}},
  \bibinfo{author}{\bibfnamefont{D.}~\bibnamefont{Bacon}},
  \bibinfo{author}{\bibfnamefont{J.}~\bibnamefont{Kempe}},
  \bibinfo{author}{\bibfnamefont{G.}~\bibnamefont{Burkard}}, \bibnamefont{and}
  \bibinfo{author}{\bibfnamefont{K.~B.} \bibnamefont{Whaley}},
  \bibinfo{journal}{Nature} \textbf{\bibinfo{volume}{408}},
  \bibinfo{pages}{339} (\bibinfo{year}{2000}).

\bibitem[{\citenamefont{Son et~al.}(2003)\citenamefont{Son, Oh, Jeong, Ahn,
  Jun, Hwang, Oh, and Engel}}]{son}
\bibinfo{author}{\bibfnamefont{M.~H.} \bibnamefont{Son}},
  \bibinfo{author}{\bibfnamefont{J.~H.} \bibnamefont{Oh}},
  \bibinfo{author}{\bibfnamefont{D.~Y.} \bibnamefont{Jeong}},
  \bibinfo{author}{\bibfnamefont{D.}~\bibnamefont{Ahn}},
  \bibinfo{author}{\bibfnamefont{M.~S.} \bibnamefont{Jun}},
  \bibinfo{author}{\bibfnamefont{S.~W.} \bibnamefont{Hwang}},
  \bibinfo{author}{\bibfnamefont{J.~E.} \bibnamefont{Oh}}, \bibnamefont{and}
  \bibinfo{author}{\bibfnamefont{L.~W.} \bibnamefont{Engel}},
  \bibinfo{journal}{Appl. Phys. Lett.} \textbf{\bibinfo{volume}{82}},
  \bibinfo{pages}{1230} (\bibinfo{year}{2003}).

\bibitem[{\citenamefont{M.Bayer et~al.}(2001)\citenamefont{M.Bayer, Hawrylak,
  Hinzer, Fafard, Korkusinski, Wasilewski, Stern, and Forchel}}]{bayer}
\bibinfo{author}{\bibnamefont{M.Bayer}},
  \bibinfo{author}{\bibfnamefont{P.}~\bibnamefont{Hawrylak}},
  \bibinfo{author}{\bibfnamefont{K.}~\bibnamefont{Hinzer}},
  \bibinfo{author}{\bibfnamefont{S.}~\bibnamefont{Fafard}},
  \bibinfo{author}{\bibfnamefont{M.}~\bibnamefont{Korkusinski}},
  \bibinfo{author}{\bibfnamefont{Z.~R.} \bibnamefont{Wasilewski}},
  \bibinfo{author}{\bibfnamefont{O.}~\bibnamefont{Stern}}, \bibnamefont{and}
  \bibinfo{author}{\bibfnamefont{A.}~\bibnamefont{Forchel}},
  \bibinfo{journal}{Science} \textbf{\bibinfo{volume}{291}},
  \bibinfo{pages}{451} (\bibinfo{year}{2001}).

\bibitem[{\citenamefont{Phillips}(1973)}]{phillips}
\bibinfo{author}{\bibfnamefont{J.~C.} \bibnamefont{Phillips}},
  \emph{\bibinfo{title}{Bonds and bands in semiconductors}}
  (\bibinfo{publisher}{Academic Press}, \bibinfo{year}{1973}).

\bibitem[{\citenamefont{Paget et~al.}(1977)\citenamefont{Paget, Lampel,
  Sapoval, and Safarov}}]{paget}
\bibinfo{author}{\bibfnamefont{D.}~\bibnamefont{Paget}},
  \bibinfo{author}{\bibfnamefont{G.}~\bibnamefont{Lampel}},
  \bibinfo{author}{\bibfnamefont{B.}~\bibnamefont{Sapoval}}, \bibnamefont{and}
  \bibinfo{author}{\bibfnamefont{V.~I.} \bibnamefont{Safarov}},
  \bibinfo{journal}{Phys. Rev. B} \textbf{\bibinfo{volume}{15}},
  \bibinfo{pages}{5780} (\bibinfo{year}{1977}).

\bibitem[{\citenamefont{bennett et~al.}(1970)\citenamefont{bennett, Watson, and
  Carter}}]{bennett}
\bibinfo{author}{\bibfnamefont{L.~M.} \bibnamefont{bennett}},
  \bibinfo{author}{\bibfnamefont{R.~E.} \bibnamefont{Watson}},
  \bibnamefont{and} \bibinfo{author}{\bibfnamefont{G.~C.}
  \bibnamefont{Carter}}, \bibinfo{journal}{J. Res. Natl. Bur. Stand. A}
  \textbf{\bibinfo{volume}{4}}, \bibinfo{pages}{74} (\bibinfo{year}{1970}).

\bibitem[{\citenamefont{Clementi and Raimondi}(1963)}]{clementi-raimondi}
\bibinfo{author}{\bibfnamefont{E.}~\bibnamefont{Clementi}} \bibnamefont{and}
  \bibinfo{author}{\bibfnamefont{D.~L.} \bibnamefont{Raimondi}},
  \bibinfo{journal}{J. Chem. Phys.} \textbf{\bibinfo{volume}{38}},
  \bibinfo{pages}{2686} (\bibinfo{year}{1963}).

\bibitem[{\citenamefont{Clementi et~al.}(1967)\citenamefont{Clementi, Raimondi,
  and Reinhardt}}]{clementi-raimondi-reinhardt}
\bibinfo{author}{\bibfnamefont{E.}~\bibnamefont{Clementi}},
  \bibinfo{author}{\bibfnamefont{D.~L.} \bibnamefont{Raimondi}},
  \bibnamefont{and} \bibinfo{author}{\bibfnamefont{W.~P.}
  \bibnamefont{Reinhardt}}, \bibinfo{journal}{J. Chem. Phys.}
  \textbf{\bibinfo{volume}{47}}, \bibinfo{pages}{1300} (\bibinfo{year}{1967}).

\end{thebibliography}
\end{document}